\documentclass[5p]{elsarticle}

\usepackage{bm}
\usepackage{amsmath}
\usepackage{amssymb}
\usepackage[colorlinks=true, allcolors=blue]{hyperref}
\usepackage{lineno}

\journal{Journal of Magnetism and Magnetic Materials}

\begin{document}

\begin{frontmatter}

\title{Equilibrium properties of magnetic filament suspensions}

\author[icmm,psu]{Andrey A. Kuznetsov\corref{cor}}
\ead{kuznetsov.a@icmm.ru}
\cortext[cor]{Corresponding author}

\address[icmm]{Institute of Continuous Media Mechanics UB RAS, Ac. Koroleva st. 1, 614013, Perm, Russia}
\address[psu]{Perm State University, Bukireva st. 15, 614990, Perm, Russia}

\begin{abstract}
Langevin dynamics is used to study equilibrium properties 
of the suspension of magnetic filaments (chains of nanoparticles permanently crosslinked with polymers).
It is shown that the filament suspension generally has a larger magnetic 
susceptibility than the system of unlinked nanoparticles with the same average particle concentration.
However, actual susceptibility gain strongly depends 
on length and flexibility of filaments.
It is also shown that in a strong gravitational (centrifugal) field
sedimentation profiles of filaments are less homogeneous
than that of unlinked particles. The spatial distribution
of filaments weakly depends on the intensity of interparticle
dipole-dipole interactions. 
\end{abstract}


\end{frontmatter}


\section{Introduction}

A magnetic filament is a chain of magnetic micro- or nanoparticles 
permanently connected by polymer linkers~\cite{cebers2005flexible,cerda2013phase}.
Such filaments have numerous potential applications.
Chains of magnetic particles 
can be used as self-propelling devices 
for drug and cargo delivery, 
as magnetically controlled
microfluidic mixers and as micromechanical 
sensors~\cite{goubault2003flexible,dreyfus2005microscopic,wang2011magnetic,cebers2016flexible}.
In Ref.~\cite{sanchez2015effect}, a 
suspension of nanosized magnetic filaments was proposed
as an improved substitute for conventional ferrofluids,
i.e. colloidal dispersions of unlinked monodomain nanoparticles
in a nonmagnetic carrier liquid.
Permanent chains are expected to have a strong response to magnetic
field and a high resistance to shear stress,
which can be a benefit in ferrofluid applications.
Much effort has been recently devoted to the analytical 
and numerical study of suspended nanosized filaments 
in the limit of infinite dilution,
when interaction between individual chains can be neglected~\cite{cerda2013phase,sanchez2015effect,cerda2016flexible}.
In Ref.~\cite{novak2017self}, a pair of interacting filaments was considered.
However, large ensembles of interacting filaments,
which are a closer approximation for filament-based ferrofluids,
have not been thoroughly studied yet.    
In the present work, equilibrium properties of
moderately concentrated filament suspensions
are numerically investigated via the Langevin dynamics simulation method.
 
\section{Model and simulation method}

The simulated system consists of $N_f$ filaments
and each filament consists of equal number of particles
$l_f$. So, the total number of particles in the system is $N_p = N_f \cdot l_f$.
Particles are uniformly magnetized spheres of equal diameter $d$
and with magnetic moments of equal constant magnitude $\mu$.
The movement of the $i$th particle
in the carrier liquid obeys the Langevin equations
\begin{linenomath}
\begin{eqnarray}
\dot{\bm{v}}^*_i = -\partial U^*_i / \partial \bm{r}^*_i - \gamma^{*T} \bm{v}^*_i + \bm{\eta}^{*T}_i, \label{langt} \\ 
J^* \dot{\bm{\omega}}^*_i = -\hat{\bm{\mu}}_i \times \partial U^*_i / \partial \hat{\bm{\mu}}_i - \gamma^{*R} \bm{\omega}^*_i + \bm{\eta}^{*R}_i, \label{langr}
\end{eqnarray}
\end{linenomath}
where asterisk denotes reduced quantities, 
$d$ is used as a unit of length,
particle mass $m$ -- as a unit of mass
and the thermal energy $k_B T$ -- as a unit of energy.
Thus, $\bm{v}^*_i = \bm{v}_i \sqrt{m / k_B T}$ 
and $\bm{\omega}^*_i = \bm{\omega}_i \sqrt{m d^2 / k_B T}$ 
are the reduced linear and angular velocities, correspondingly,
$\bm{r}^*_i = \bm{r}_i / d$ is the reduced particle position,
$\hat{\bm{\mu}}_i = \bm{\mu}^*_i / \mu^*  = \bm{\mu}_i / \mu$ is the unit vector along the particle magnetic moment,
$\mu^* = \mu \sqrt{\mu_0/ 4\pi d^3 k_B T}$ is the reduced magnetic moment,
$\mu_0$ is the magnetic constant,
$U^*_i = U_i / k_B T$ is the reduced particle potential energy,
$J^* = J / m d^2$ is the reduced moment of inertia,
$\gamma^{*T} = \gamma^{T} \sqrt{d^2 / m k_B T}$ and
$\gamma^{*R} = \gamma^{R} \sqrt{1 / d^2 m k_B T}$ 
are the reduced translational and rotational friction coefficients, 
$\bm{\eta}^{*T}_i$ and $\bm{\eta}^{*R}_i$ are the random Gaussian force and torque,
which have zero mean values and satisfy the 
standard fluctuation-dissipation relationship
\begin{linenomath}
\begin{equation}
\langle \eta^{*T(R)}_{i \alpha}(t^*_1) \eta^{*T(R)}_{j \beta}(t^*_2) \rangle = 2 \gamma^{*T(R)}\delta_{\alpha \beta} \delta_{ij}\delta^*(t^*_1 - t^*_2), 	
\end{equation} 
\end{linenomath}
the reduced time is $t^* = t \sqrt{k_B T / m d^2}$.

The interaction energy of two arbitrary particles
$i$ and $j$ consists of the steric repulsion energy $u_{sr}(i,j)$
and the dipole-dipole interaction energy $u_{dd}(i,j)$:
\begin{linenomath}
\begin{eqnarray}  
u_{sr}(i,j) =
\begin{cases}
u_{LJ}(r_{ij}) - u_{LJ}(r_{cut}), & r_{ij} < r_{cut} \\
0, & r_{ij} \geq r_{cut},
\end{cases}, \label{uwca} \\
u_{LJ}(r) = 4 \varepsilon\left[\left(\frac{d}{r}\right)^{12} - \left(\frac{d}{r}\right)^{6}\right],\\
u_{dd}(i,j) = 
\frac{\mu_0}{4\pi}
\left[ 
	\frac{\bm{\mu}_i \cdot \bm{\mu}_j}{r^{3}_{ij}} - 
	\frac{3(\bm{\mu}_i \cdot \bm{r}_{ij})(\bm{\mu}_j \cdot \bm{r}_{ij})}{r^{5}_{ij}}
\right],
\end{eqnarray}
\end{linenomath}
where $u_{LJ}$ is the Lennard-Jones potential,
$r_{cut} = 2^{1/6}d$ is the cutoff radius~\cite{allen1987}.
Additionally, some bonding interaction between adjacent particles
in the filament must be imposed to preserve
filament's chain-like structure.
For this purpose, the phenomenological potential introduced in 
Refs.~\cite{cerda2013phase,sanchez2015effect} is used:
\begin{linenomath}
\begin{equation}
u_{bond}(i+1,i) = \frac{K}{2}\left[\bm{r}_{i+1,i} - (\hat{\bm{\mu}}_{i+1} + \hat{\bm{\mu}}_i)\frac{d}{2}\right]^2. \label{uspring}
\end{equation} 
\end{linenomath}
The potential mimics the harmonic spring that links
opposite poles of two consequent particles in the filament.
It is not only forbids neighbors to move too far from each other,
but also promotes an alignment of their magnetic moments
(it is implied that the magnetic anisotropy of particles
is large and moments are ``frozen'' within the particle bodies).
Three dimensionless energy parameters can be introduced for 
the simulated system: the steric repulsion parameter $\epsilon = \varepsilon / k_B T$,
the dipolar coupling parameter $\lambda = (\mu_0 / 4 \pi) \mu^2 /  d^3 k_B T = \mu^{*2}$
and the elastic parameter $\varkappa = Kd^2/2 k_B T$.
Using these parameters, interaction potentials (\ref{uwca})--(\ref{uspring}) can be rewritten in the reduced form:
\begin{linenomath}
\begin{eqnarray}  
u^*_{sr}(i,j) =
\begin{cases}
4 \epsilon \left( \frac{1}{r^{*12}} - \frac{1}{r^{*6}} + \frac{1}{4} \right), & r^*_{ij} < r^*_{cut}, \\
0, & r^*_{ij} \geq r^*_{cut},
\end{cases}, \label{uwca2} \\
u^*_{dd}(i,j) = 
\lambda
\left[ 
\frac{\hat{\bm{\mu}}_i \cdot \hat{\bm{\mu}}_j}{r^{*3}_{ij}} - 
\frac{3(\hat{\bm{\mu}}_i \cdot \bm{r}^{*}_{ij})(\hat{\bm{\mu}}_j \cdot \bm{r}^{*}_{ij})}{r^{*5}_{ij}}
\right], \\
u^*_{bond}(i+1,i) = \varkappa\left[\bm{r}^*_{i+1,i} - \frac{\hat{\bm{\mu}}_{i+1} + \hat{\bm{\mu}}_i}{2}\right]^2.
\end{eqnarray}
\end{linenomath}

The input parameters of the simulation are
$\varkappa$, $\lambda$, $l_f$, $N_p$ and the average particle
volume fraction $\overline{\varphi} = N_p \nu / V$,
where $\nu = \pi d^3 / 6$ is the particle volume,
$V$ is the volume of the simulation cell.
Other parameters are fixed: $\epsilon = 0.67$, $J^* = 0.1$,
$\gamma^{*R} = 3$, $\gamma^{*T} = 1$.
$\epsilon = 0.67$ ensures that at $\varkappa = 0$
and $\lambda = 0$ the system free energy coincides with the
free energy of the ensemble of hard spheres with a diameter $d$~\cite{kuznetsov2017sedimentation}.
Surely, in a more general case, when $\lambda > 0$ or $\varkappa > 0$,
equilibrium properties of filaments will still depend on the particular
choice of the steric repulsion potential.
The usage of the ``soft sphere'' potential Eq.~(\ref{uwca2}) in the present study 
is physically grounded since in most cases particles are stabilized by soft polymer shells~\cite{sanchez2015effect}. 
Langevin equations are integrated using the modified leapfrog 
algorithm by Gr{\o}nbech-Jensen and Farago~\cite{Gronbech-Jensen2014}.

\section{Results} \label{sec:res}

\subsection{Filament configuration at infinite dilution: effect of the elastic parameter}

\begin{figure}
	\centering
	\includegraphics[scale=.9]{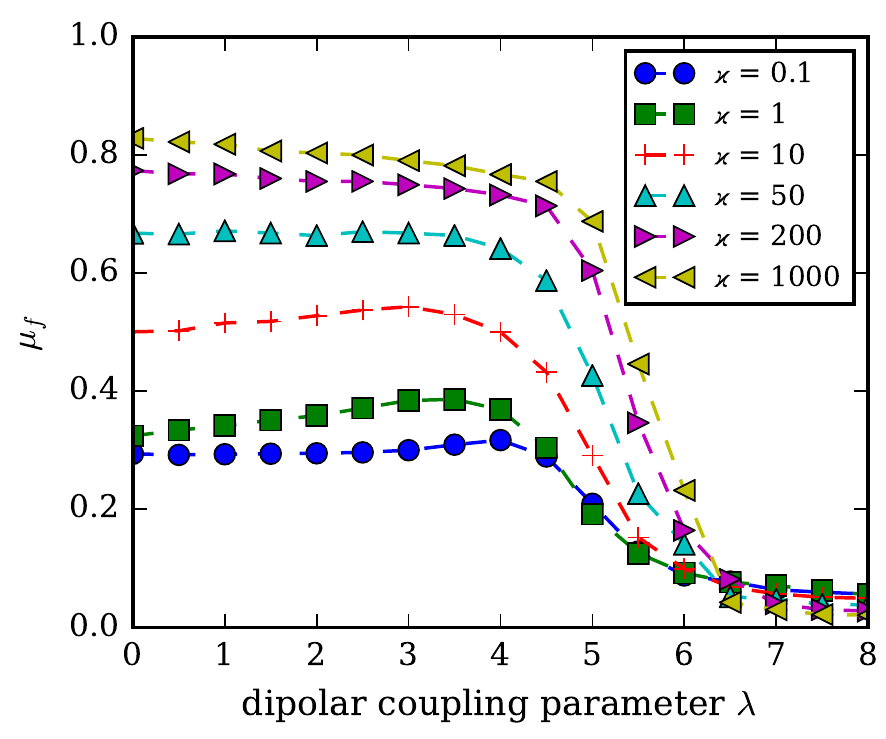}
	\caption{\label{fig:elastic_effect} 
		Average normalized magnetic moment of the isolated filament
		vs. the dipolar coupling parameter at different values of the elastic parameter $\varkappa$. $N_f = 1$, $l_f = 10$.}
\end{figure}

Auxiliary simulations were conducted in order to 
analyze the effect of the elastic parameter $\varkappa$ on the behavior
of the considered filament model.
A single isolated ten-particle filament ($N_f = 1$, $l_f = 10$)
was simulated in the absence of external force fields 
and under open boundary conditions (similarly to Refs.~\cite{cerda2013phase,sanchez2015effect}).
Normalized average magnetic moment of the filament 
\begin{linenomath}
\begin{equation}
\mu_f = \frac{1}{ l_f \mu}\left\langle \left|\sum_{i = 1}^{l_f} \bm{\mu}_i \right|\right\rangle
\end{equation}
\end{linenomath}
was calculated for different values of $\lambda$ and $\varkappa$.
Results are shown in Fig.~\ref{fig:elastic_effect}.
Let's first consider $\lambda = 0$. 
In this limiting case, dipole-dipole 
interactions between cross-linked nanoparticles are negligible. 
It is possible, for example, when particles have thick polymer shells with a characteristic diameter $d \gg d_c = [(\mu_0/4\pi)\mu^2/k_B T]^{1/3}$.
At $\varkappa = 0.1$ $\mu_f \simeq 1/\sqrt{l_f} \simeq 0.3$,
which coincides with the average magnetic moment of $l_f$
independently fluctuating dipoles.
\begin{figure}	
	\centering
	\includegraphics[scale=0.8]{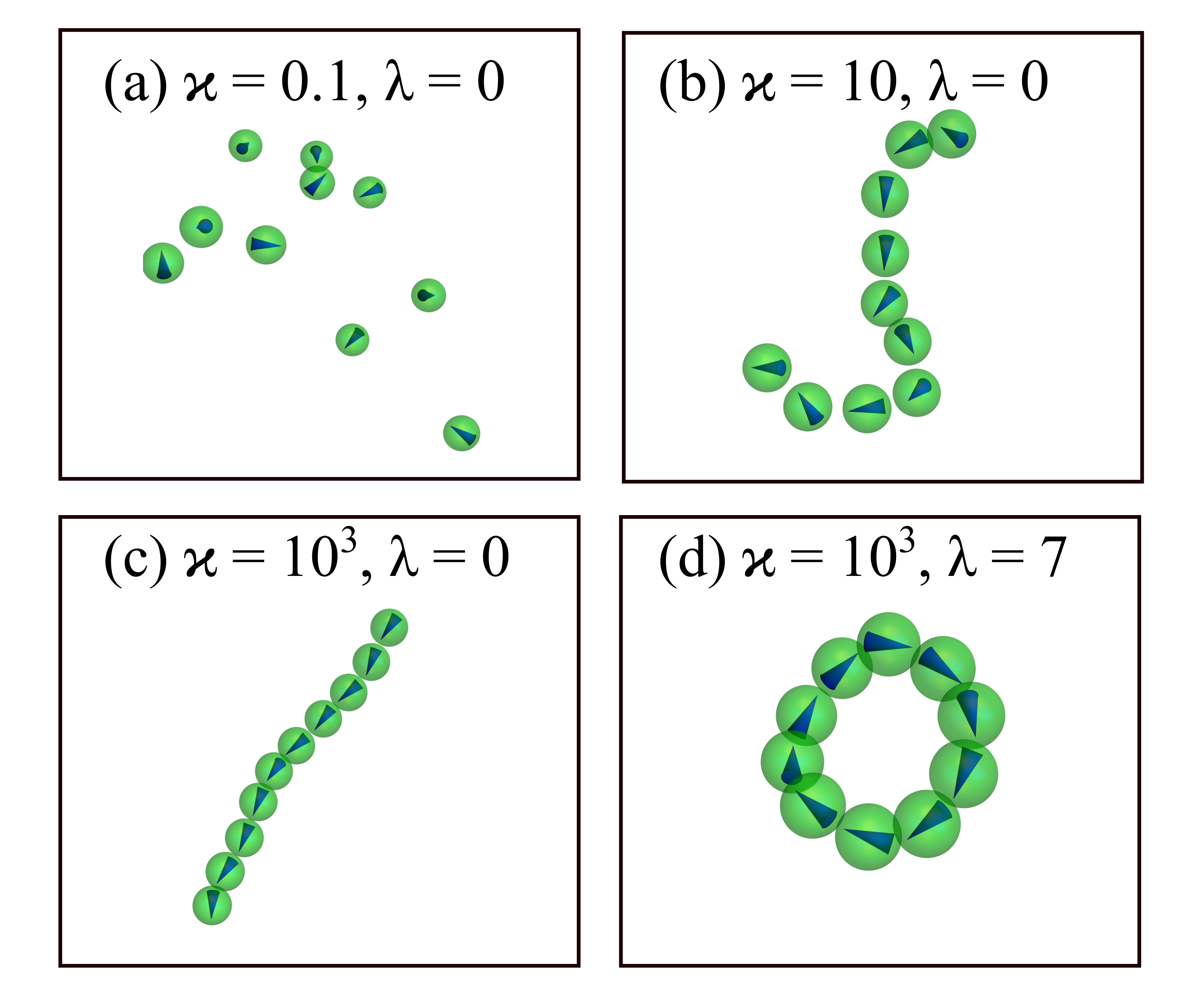}
	\caption{\label{fig:snaps_single} 
		3D snapshots of the isolated filament
		at different values of elastic ($\varkappa$) and dipolar coupling ($\lambda$) parameters. $N_f = 1$, $l_f = 10$.}
\end{figure}
As seen in Fig.~\ref{fig:snaps_single}(a),
particles are chaotically moving in the proximity of each other
and do not resemble a chain.
An increase of $\varkappa$ by two orders of magnitude 
increases magnetic moment only up to $\mu_f \simeq 0.5$.
At this point the system obtains the chain-like structure, 
but the chain is rather flexible (see Fig.~\ref{fig:snaps_single}(b)).
At $\varkappa = 10^3$ (Fig.~\ref{fig:snaps_single}(c)) $\mu_f \gtrsim 0.8$,
which is close to the 
magnetic moment of a rigid rodlike chain ($\mu_f = 1$).
At $\lambda \le 4$ the magnetic moment only slightly depends on the dipolar parameter,
but at $4 \lesssim \lambda \lesssim 6$ it rapidly decreases.
At $\lambda > 6$ filaments are in the stable ring configuration with $\mu_f < 0.1$ (Fig.~\ref{fig:snaps_single}(d)).
The ring formation occurs regardless the elastic parameter.
In what follows only systems with $\lambda < 4$ will be considered
since rings are known to weaken the magnetic response of ferrofluids~\cite{kantorovich2013nonmonotonic}.


\subsection{Initial magnetic susceptibility of the filament suspension}

\begin{figure}
	\centering
	\includegraphics[width=0.3\textwidth]{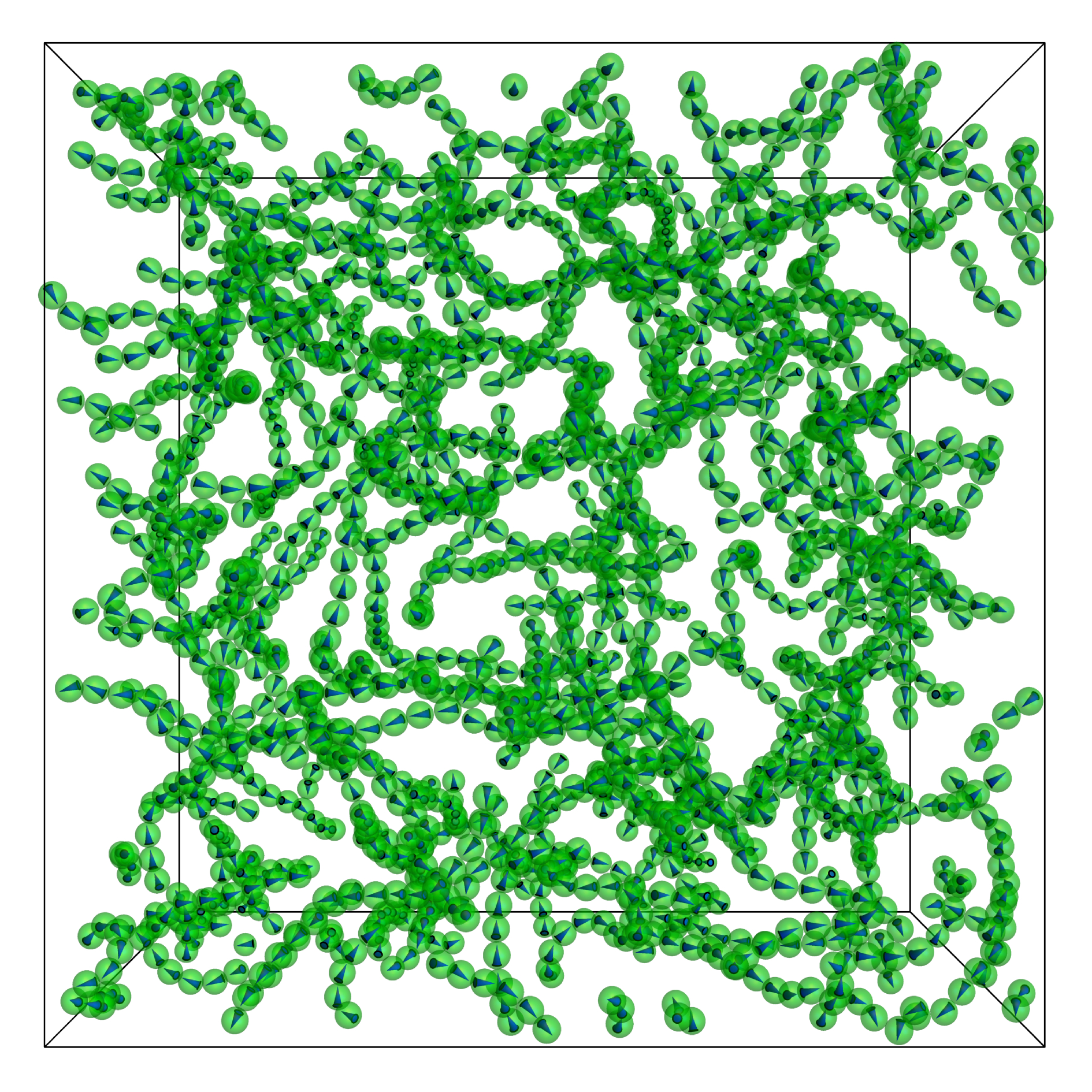}
	\caption{\label{fig:snap}
		Snapshot of the filament ensemble inside the cubic box
		with 3D periodic boundary conditions.
		$N_p = 2000$, $\overline{\varphi} = 0.025$,
		$l_f = 10$, $\lambda = 2$, $\varkappa = 10^3$.}
\end{figure}

\begin{figure}
	\centering
	\includegraphics[scale=0.9]{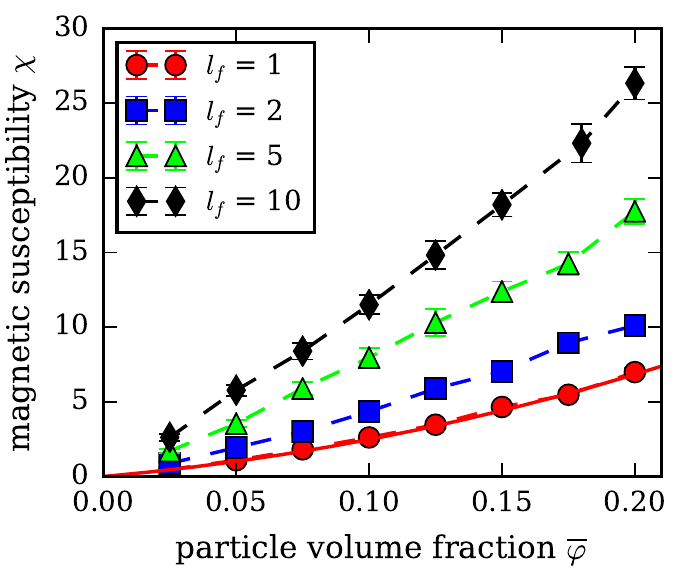}
	\caption{\label{fig:susc} 
		Initial susceptibility of the filament ensemble
		vs. the average particle volume fraction at different filament lengths.
		Solid curve is from Eq.~(\ref{eq:mmef2}).
		$N_p = 2000$, $\lambda = 2$, $\varkappa = 10^3$.}
\end{figure}

\begin{figure}
	\centering
	\includegraphics[scale=0.8]{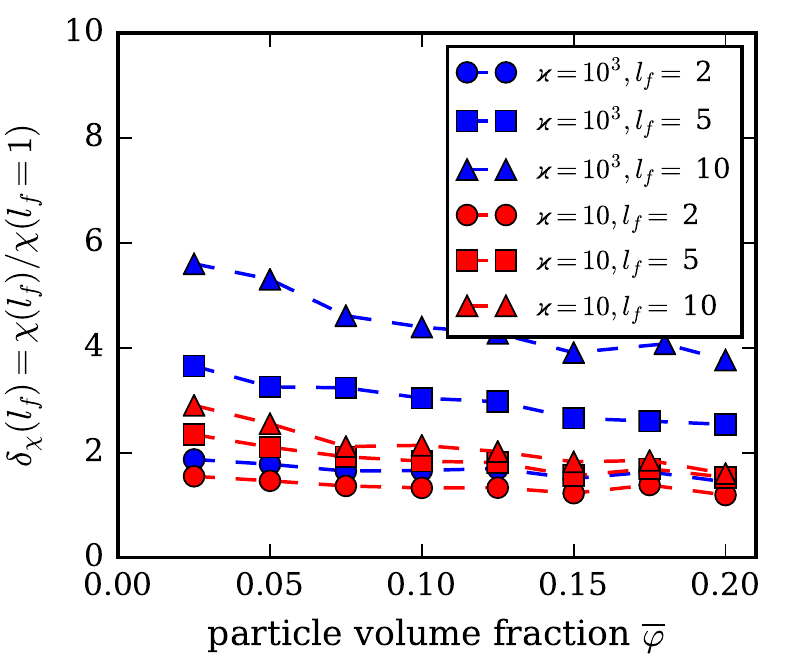}
	\caption{\label{fig:the_quantity} 
		Susceptibility gain due to replacement of individual particles
		in the suspension with filaments vs. the average particle volume fraction. $N_p = 2000$, $\lambda = 2$.}
\end{figure}

To calculate the magnetic susceptibility
of the filament suspension, the system of $N_p = 2000$
particles in the cubic cell with 3D periodic boundary 
conditions was simulated.   
Standard Ewald summation technique with ``conducting'' boundary conditions
was used to calculate dipole-dipole interactions in the periodic system~\cite{allen1987}.
External fields were absent.
Figure~\ref{fig:snap} shows an example of the system equilibrium configuration.
Susceptibility was calculated as follows~\cite{de1986computer}:
\begin{linenomath}
\begin{equation} 
\chi = \chi_L \left\langle \left(\sum_{i = 1}^{N_p} \bm{\mu}_i\right)^2 \right\rangle \frac{1}{\mu^2 N_p},
\end{equation}
\end{linenomath}
where $\chi_L = 8 \lambda \overline{\varphi}$ is the Langevin susceptibility.
In Fig.~\ref{fig:susc} concentration dependencies of $\chi$
are given for $\lambda = 2$, $\varkappa = 10^3$ and different filament sizes $l_f$.
For $l_f = 1$, i.e. when particles are not forced to form chains, 
results are well described by the second-order modified mean-field model~\cite{ivanov2001magnetic}
\begin{linenomath}
\begin{equation} \label{eq:mmef2}
\chi = \chi_L (1 + \chi_L/3 + \chi^2_L/144).
\end{equation}  
\end{linenomath}
For $l_f > 1$ susceptibility is expectedly larger.
Figure~\ref{fig:the_quantity} shows concentration dependencies of the
quantity $\delta_{\chi}(l_f) = \chi(l_f)/\chi(l_f = 1)$.
Basically, this quantity indicates how the replacement of 
individual colloidal particles with filaments 
increases the suspension magnetic response,
if the amount of magnetic phase, its average concentration 
and the dipolar coupling parameter remain unchanged. 
It is seen that for flexible filaments ($\varkappa = 10$)
$\delta_{\chi}$ weakly increases with the filament size,
in a broad range of concentrations $\delta_{\chi} \lesssim 2$
regardless $l_f$ value.
For stiff chains ($\varkappa = 10^3$) 
a much stronger increase of the susceptibility can be achieved.
Though even in this case $\delta_{\chi} (l_f) < l_f$.
It also seems that for fixed $\lambda$ and $l_f$ 
the quantity $\delta_{\chi}$ decreases with the growth
of the average particle concentration.
To clarify this effect, further investigations are required. 


\subsection{Sedimentation stability of the filament suspension}

From an applied point of view, one of the most important 
characteristics of ferrofluids is their ability 
to maintain spatial homogeneity under the action
of applied force fields such as gravitational
or gradient magnetic ones~\cite{pshenichnikov2015magnetic}.
To evaluate the stability of the filament-based 
ferrofluid, the process of filaments' sedimentation
in a strong gravitational (centrifugal) field was simulated.

Simulation algorithm proposed in Ref.~\cite{kuznetsov2017sedimentation} was used.
Filaments were placed inside the elongated rectangular cavity of the height $L_z$,
gravitational field $\bm{g}$ was acting along the cavity long axis ($z$-axis).
Top and bottom surfaces of the cavity ($z = L_z$ and $z = 0$, correspondingly) 
were impenetrable for particles and 2D periodic boundary conditions were imposed along the $x-$ and $y-$ directions.
To take into account the presence of the gravitational field,
additional term in the $i$th particle's full energy was introduced: $u_g(i)/k_BT = z_i/L_{sed}$, 
where $L_{sed} = k_B T / \Delta \rho \nu g$ is the particle sedimentation length,
$\Delta \rho$ is the density difference between the particle and the carrier liquid.
Long-range dipole-dipole interactions were calculated using the modified Ewald
summation technique adapted for the slab geometry. Details
can be found in Ref.~\cite{klapp2002spontaneous}.
The main simulation result is the equilibrium concentration profile $\varphi = \varphi(z)$,
where $\varphi$ is the \textit{local} volume fraction of particles at the height $z$.

\begin{figure}
	\centering
	\includegraphics{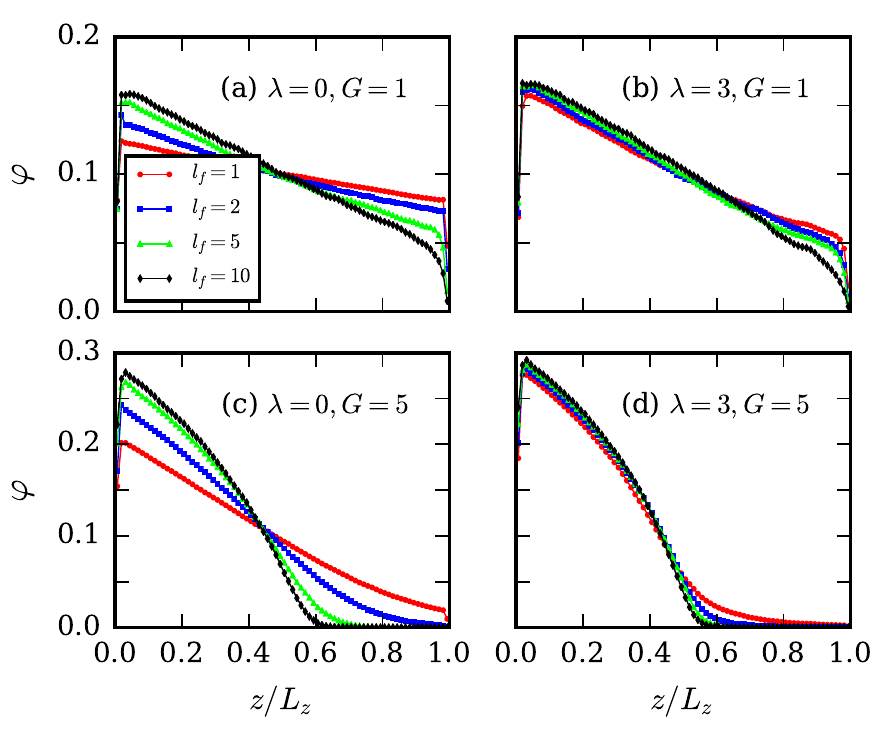}
	\caption{\label{fig:profiles}
		Sedimentation profiles of filaments with different lengths
		at different values of the dipolar coupling parameter ($\lambda$) and
		the gravitational parameter ($G = L_z/L_{sed}$).
		$N_p = 2000$, $\overline{\varphi} = 0.1$, $L_z = 80d$, $\varkappa = 10^3$.}
\end{figure}

Profiles obtained at different values of $l_f$, $\lambda$ 
and the gravitational parameter $G = L_z/L_{sed}$ are given in Fig.~\ref{fig:profiles}.
It is seen that at $\lambda = 0$ the suspension with $l_f = 10$
is substantially more inhomogeneous than the suspension with $l_f = 1$.
It simply means that the gradient diffusion coefficient of stiff
chains is lower than that of unlinked soft spheres.
In the case of ten-particle filaments,
profiles do not significantly change with the growth of $\lambda$.
On the contrary, the spatial homogeneity of unlinked particles is 
strongly affected by dipole-dipole interactions.
At $\lambda = 3$ profiles for $l_f = 1$ and $l_f = 10$ become very close.
This is probably due to the fact that 
at high $\lambda$ magnetic particles 
tend to assemble into various aggregates even without polymer linkers~\cite{kantorovich2013nonmonotonic}. 

\section{Conclusion}

It was numerically investigated how
the replacement of individual nanoparticles
with nanosized magnetic filaments will affect 
the equilibrium properties of ferrofluids.
It was shown that at the same average concentration of magnetic phase
the filament suspension have a stronger magnetic response than the standard ferrofluid.
However, to obtain a large
susceptibility gain (more than twofold), several requirements must be met.
Filaments must not only be long, they have to be rather stiff:
the characteristic energy of the polymer bonding between 
two adjacent particles in the filament
must be several orders of magnitude larger than the thermal energy. 
The replacement of particles with filaments has a negative impact 
on the suspension stability in strong gravitational (centrifugal) fields.
At high dipolar coupling parameters $\lambda$ the effect is weak: 
concentration profiles for unlinked particles (monomers) and for the system 
of ten-particle filaments are almost the same.
In both cases, magnetic phase distribution can be highly inhomogeneous.
But the inhomogeneity of monomers decreases with decreasing $\lambda$, 
whereas for long filaments the suspension remains segregated even when dipole-dipole interactions are weak.

Obtained results suggest that the replacement of individual 
particles with filaments 
will have a greater impact on the system susceptibility
if the average particle concentration is low.
It means that interchain interactions (steric and/or dipole-dipole ones) are able to reduce the advantages of magnetic filament suspensions 
over standard ferrofluids.  
The exact mechanism behind this effect is not clear.
Perhaps, a detailed study of filaments' microstructure 
at different concentrations will shed light on the problem.
Another issue that deserves a special consideration is 
the stability of filament suspensions in high-gradient magnetic fields,
which are typical for some applications of ferrofluids~\cite{pshenichnikov2015magnetic,krakov2014effect}.
These problems will be addressed in future articles.

\section{Acknowledgments}

The work was supported by the 
Foundation for Assistance to Small Innovative Enterprises in Science and Technology 
(agreement No. 8954GU/2015).


\bibliography{E:/main/library/magnetic_particles_lib}

\end{document}